\documentclass[conference]{IEEEtran}
\usepackage{cite}

\ifCLASSINFOpdf
   \usepackage[pdftex]{graphicx}
\else
\fi
\usepackage{epstopdf}

\usepackage[cmex10]{amsmath}
\usepackage{algorithm}
\usepackage{algorithmic}

\usepackage{array}

\usepackage{mdwmath}
\usepackage{mdwtab}

\usepackage{eqparbox}

\usepackage[tight,footnotesize]{subfigure}

\usepackage{eqnarray}

\usepackage[font=footnotesize]{subfig}
\usepackage{stfloats}

\usepackage{url}

\begin{document}
%
\title{Radio Link Enabler for Context-aware D2D Communication in Reuse Mode}

\author{\IEEEauthorblockN{Ji Lianghai, Andreas Weinand, Michael Karrenbauer, Hans D. Schotten}
\IEEEauthorblockA{Chair of Wireless Communication, TU Kaiserslautern\\
Email: $\lbrace$ji,weinand,karrenbauer,schotten$\rbrace$@eit.uni-kl.de}}

\IEEEoverridecommandlockouts
\IEEEpubid{\makebox[\columnwidth]{\copyright~Copyright 2017
		IEEE \hfill} \hspace{\columnsep}\makebox[\columnwidth]{ }}

\maketitle

\begin{abstract}
Device-to-Device (D2D) communication is considered as one of the key technologies for the fifth generation wireless communication system (5G) due to certain benefits provided, e.g. traffic offload and low end-to-end latency. A D2D link can reuse resource of a cellular user for its own transmission, while mutual interference in between these two links is introduced. In this paper, we propose a smart radio resource management (RRM) algorithm which enables D2D communication to reuse cellular resource, by taking into account of context information. Besides, signaling schemes with high efficiency are also given in this work to enable the proposed RRM algorithm. Simulation results demonstrate the performance improvement of the proposed scheme in terms of the overall cell capacity. 
\end{abstract}


%
\IEEEpeerreviewmaketitle

\section{Introduction}
Recently, network controlled device-to-device (D2D) communications has attracted a lot of attention in research, since it is able to efficiently complement cellular communications by taking advantage of physical proximity for improving coverage, spectral efficiency, data rates, and QoS \cite{re1}. Moreover some new services can also be enabled by D2D communication, such as direct multimedia transmission \cite{re2}\cite{re3}. The key functionality for D2D communications include \cite{re1}:
\begin{itemize}
\item Peer and Service Discovery
\item Physical Layer Procedures (encoding, signaling, data transmission and reception, etc.),
\item Radio Resource Management (RRM) (transmit power and resource allocation, data rates, etc.),
\item Interference Management (e.g. intra-cell interference is no longer negligible due to use in LTE).
\end{itemize}
As one of the important issues in D2D communications, RRM covers the concepts of D2D transmit power control, mode selection and resource allocation. Power control is the mechanism of optimally selecting transmit powers in a communication system to achieve good performance and reduce interference \cite{re24}\cite{re25}\cite{re26}. Good performance in this context may be assessed using metrics, such as link data rates, network capacity, geographic coverage and range, etc. Transmit power determines the range over which the signal can be coherently received, and is therefore crucial for determining the performance of the network (e.g. with respect to throughput, delay, and energy consumption) \cite{re20}. In order to realize the proximity, reuse and hop gains, the available spectral resource need to be allocated for D2D communications in an efficient way. Presently, there are three modes of D2D operation that can be envisioned:
\begin{itemize}
\item Reuse Mode: D2D devices directly transmit the data by reusing some resources of the cellular network. The spectrum reuse can be either in uplink or downlink communications.
\item Dedicated Mode: The cellular network dedicated a portion of resources for D2D devices for their direct communications.
\item Cellular Mode: D2D traffic is relayed through eNB’s in the traditional way.
\end{itemize}
Some solutions exclude the coexistence of the D2D and cellular communications using the same spectrum resource and therefore the dedicated mode is considered \cite{re1}\cite{re23}\cite{re30}. However, in order to improve the efficiency of valuable spectrum resources, the reuse mode for D2D is also investigated where the same spectrum resource is shared between D2D and cellular users \cite{re1}\cite{re30}\cite{re27}\cite{re28}\cite{re29}. The reuse mode becomes important when available frequency range is considered as precious and it is essential to reuse the same spectrum resource.\\
In this paper, it is assumed that D2D communication operates in reuse mode where uplink resource of cellular users (UEs) are reused by D2D links. In Sect.~\ref{SM}, the system model is introduced and the interference issues that are raised by D2D in reuse mode are illustrated. After, in Sect.~\ref{challenge}, we describe the challenge of collecting certain context information which are used by RRM algorithm to control mutual interference in between cellular links and D2D links in reuse mode. An alternative RRM algorithm with a reasonable signaling effort is also proposed in Sect.~\ref{challenge}. Further, signaling schemes used to enable the proposed RRM algorithm are illustrated in Sect.~\ref{signaling}. The simulation models and numerical results shown in Sect.~\ref{NR} demonstrate system performance of our proposed solutions. Finally, we draw a conclusion of our work in Sect.~\ref{con}.
%
\section{System description}\label{SM}
\begin{figure}[!t]
\centering
\includegraphics[width=3.5in]{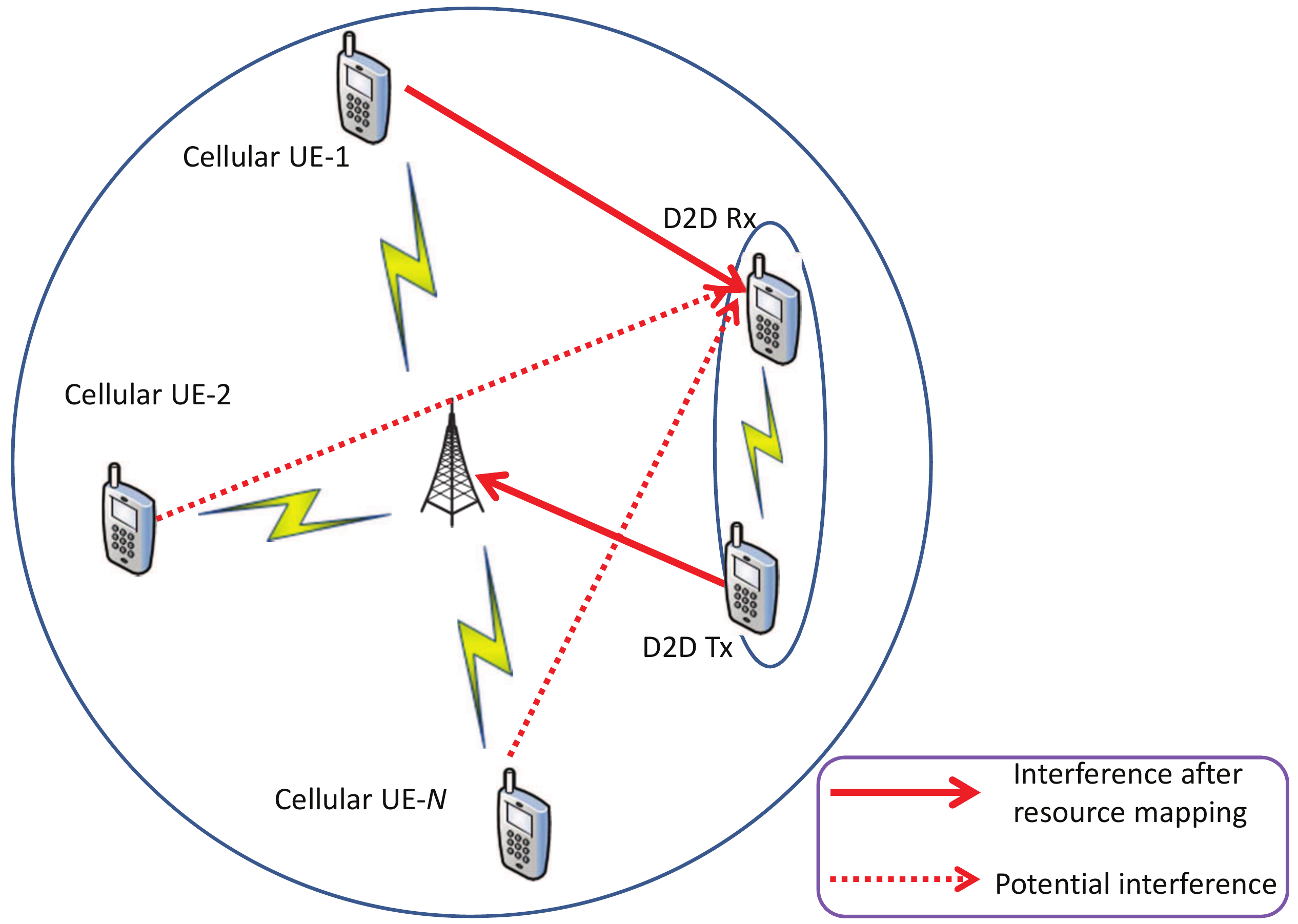}
\caption{Interference scenario: D2D communication in reuse mode}
\centering
\label{D2D}
\end{figure}

We consider a scenario where mobile users are divided into two categories, cellular users and D2D users. Cellular users request communication service directly from a base station (BS). In comparison, two nearby D2D users can form a D2D pair and perform a local communication service in order to exchange data with each other. Since D2D communication in reuse mode is assumed in our work, there are no dedicated resources available for data transmission of D2D communication. Hence a D2D pair can only reuse uplink resource block of a cellular user. Further, it is assumed that the uplink resource of cellular user can be reused by maximally one D2D link in one cell sector.\\
In Fig.~\ref{D2D}, D2D communication in reuse mode tries to reuse resource blocks that are already occupied by cellular uplinks. With total awareness of channel information, the BS can assign one cellular resource block to a D2D pair with certain optimization consideration, i.e. to enable more D2D links in total or maximize the overall system capacity. In order to fulfil the QoS requirement of both cellular and D2D links, when one cellular uplink and one D2D links are assigned with the same time-frequency resource, the received signal quality should be taken into account by BS. Therefore, the $m^{th}$ D2D pair can reuse resource block of the $n^{th}$ cellular UE, if and only if the SINR values of both the D2D link and cellular link are higher than the target values $SINR_{target}^{D2D}$ and $SINR_{target}^{cell}$, as
\begin{equation}\label{con1}
SINR_{(m,n)}^{D2D}>=SINR^{D2D}_{target},
\end{equation} 
and
\begin{equation}\label{con2}
SINR_{(m,n)}^{cell}>=SINR^{cell}_{target}.
\end{equation}
The $SINR_{(m,n)}^{D2D}$ and $SINR_{(m,n)}^{cell}$ values are calculated as:
\begin{equation}\label{sinrD2D_c}
SINR_{(m,n)}^{D2D}=\frac{h_m^{D2D}P_T^{D2D}}{h_{(m,n)}P_T^{cell}+\sigma^{2}},
\end{equation}
\begin{equation}\label{sinrCell_c}
SINR_{(m,n)}^{cell}=\frac{h_n^{cell}P_T^{cell}}{h_{(BS,m)}P_T^{D2D}+\sigma^{2}},
\end{equation}
$P_T^{D2D}$ and $P_T^{cell}$ denote the transmission powers of D2D and cellular transmitters, respectively. $h_m^{D2D}$ represents the channel gain between transmitter and receiver of the $m^{th}$ D2D pair, $h_n^{cell}$ the channel gain between the $n^{th}$ cellular UE and BS. $h_{(m,n)}$ represents the channel gain from the $n^{th}$ cellular UE to $m^{th}$ D2D receiver (Rx), $h_{(BS,m)}$ the channel gain from the $m^{th}$ D2D transmitter (Tx) to BS. One channel gain value includes the effects from pathloss model, shadowing and antenna gain. $\sigma^2$  is used here to denote the thermal noise power, including influence of receiver noise figure.\\
A feasibility function can be constructed to show whether a D2D link and a cellular link can be assigned with the same resource while the SINR requirements are fulfilled for both the cellular and D2D links, as:
\begin{equation}\label{defF}
f(m,n)=\begin{cases}
1, & \text{if Eq.~(\ref{con1}) and Eq.~(\ref{con2}) are fulfilled};\\
0, & \text{else}.
\end{cases}
\end{equation}
The feasibility of assigning same resource to both cellular and D2D links should be taken into account by BS with a smart radio resource management algorithm (RRM).
\section{Challenge and the proposed solution}\label{challenge}
It can be seen from Eq.~(\ref{sinrD2D_c}) and Eq.~(\ref{sinrCell_c}) that channel state information (CSI) of overall cellular and D2D links are required to be available at BS, in order to support a smart RRM algorithm for D2D communication. In a single cell scenario with $M$ cellular UEs and $N$ D2D pairs, a total awareness of CSI means the availability of following information at BS:
\begin{itemize}
\item $M$ cellular link channel gain values (each for one link between uplink cellular UE and BS).
\item $N$ D2D link channel gain values (each for one link between D2D Tx and Rx).
\item $N$ channel gain values to characterize interference links for cellular UEs (each for one link between D2D Tx and BS). 
\item $M\times N$ channel gain values to characterize interference links for D2D links (each for link between one uplink cellular UE and one D2D Rx).
\end{itemize}
As analyzed above, the $M\times N$ channel gain values characterizing interference links for D2D links are used to calculate the $SINR_{(m,n)}^{D2D}$ values as shown in Eq.~\ref{sinrD2D_c}. These information are collected in a cumbersome manner and will cause a large signaling overhead.\\
\begin{figure*}[!b]
\centering
\includegraphics[width=6in]{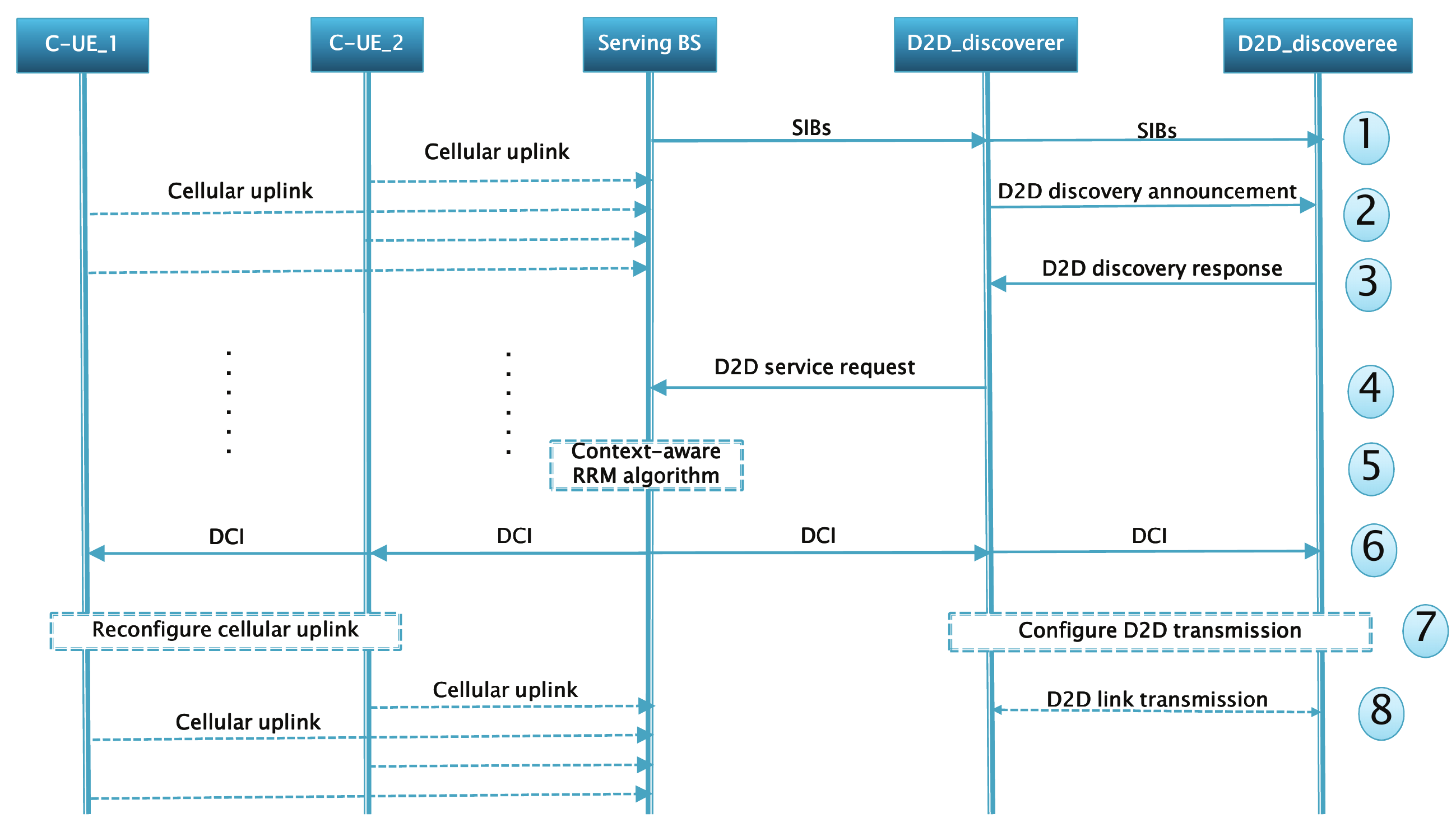}
\caption{Signaling scheme in single cell}
\label{SSS}
\end{figure*}
In order to reduce signaling overhead and achieve an efficient signaling scheme, we propose a smart RRM scheme where UEs’ location information are exploited by BS to decide whether one D2D link can reuse resource of a cellular UE. In this scheme, interference from a cellular link to a D2D receiver is considered to be under control if following equation fulfills:
\begin{equation}\label{DisRatio}
\frac{d_{(m,n)}}{d_m}\geq \gamma_{(SINR_{target}^{D2D},d_m,\cdots)};
\end{equation}
$d_m$ is the distance between two UEs of the $m^{th}$ D2D pair, and $d_{(m,n)}$ is the distance from the $n^{th}$ cellular UE to the receiver of the $m^{th}$ D2D pair. $\gamma_{(SINR_{target}^{D2D},d_m,\cdots)}$ is a threshold value which can be a function of $SINR_{target}^{D2D}$, $d_m$ and other useful context information. With this alternative way of estimating channel situation, the feasibility function in Eq.~(\ref{defF}) can be rewritten, as
\begin{equation}\label{estF}
\tilde{f}_{(m,n)}=\begin{cases}
1, & \text{if Eq.~(\ref{con2}) and Eq.~(\ref{DisRatio}) are fulfilled};\\
0, & \text{else}.
\end{cases}
\end{equation}
\section{Signaling schemes to support D2D communication}\label{signaling}
Context information have been proposed to be used in literature \cite{my1}\cite{my2} to control mutual interference. However, there was no clear illustration regarding the corresponding signaling scheme design in control plane. Without the support of control plane design, any RRM algorithms designed to offload cellular traffic to D2D links can not be enabled. Thus, we show our proposed signaling schemes which can support our context-aware RRM algorithm with good efficiency. 
\subsection{Signaling scheme in single cell}
Fig.~\ref{SSS} shows the proposed signaling scheme where two D2D ends are served by one BS. The D2D service authorization process is not considered here since it is not the main topic for this work. In this case, several cellular users are served by one base station already with dedicated resource. And D2D link tries to request assignment of resource from BS. Dashed lines are used in this figure to represent cellular uplink transmission sessions. The eight steps involved in the shown signaling diagram are illustrated with more details in following, as
\begin{enumerate}
\item[1)] Users exploiting D2D service receive system information blocks (SIBs) broadcasted by BS and dedicated to D2D operation. In these SIBs, system information for supporting D2D discovery and communication processes are provided, i.e. configuration information of resource pools used for D2D link transmission \cite{36331}. 
\item[2)] One D2D discoverer UE transmits one discovery message to potential D2D discoveree UEs by exploiting either ProSe direct discovery model A or model B, as defined in 3GPP \cite{23303}. In model A, D2D discoverer transmits a discovery message by announcing "I AM HERE". And in model B, a discovery message "WHO IS THERE?" or "ARE YOU THERE?" will be transmitted. Along with the discovery message, application information and reference signals for D2D channel estimation are also transmitted.
\item[3)] A D2D discovery receiver who receives the discovery message successfully checks whether it is permitted to form a D2D link with the transmitter and afterwards announces a response message to indicate its availability.  If the discoveree transmits an acknowledgement message to D2D discoverer, certain context information will also be conveyed in the message, i.e. D2D discoveree position and velocity and the measured channel gain information for both D2D link and cellular link between serving BS and the D2D discoveree.
\item[4)]  Upon receiving of an acknowledgement message, D2D discoverer sends a service request to the serving BS, together with certain context information, i.e. channel gain information of both D2D and cellular links, QoS requirements of the D2D link, user positions and velocities of both D2D ends, service priority. The D2D channel measurement is performed in step 2). Besides, channel measurement of cellular links is performed in the same way as in legacy LTE-A system by measuring the cell specific reference signals (CSRS).
\item[5)] BS analyses the gathered context information and performs its context-aware RRM algorithm. 
\item[6)] If RRM algorithm admits corresponding D2D link to reuse resource of uplink cellular users, a resource configuration message will be sent by BS to the D2D ends in downlink control information (DCI). Meanwhile, cellular links with resource being reused may also be reconfigured, in order to reduce impact from D2D link.
\item[7)] After receiving the configuration information, D2D link configures its transmission with the allocated resource.
\item[8)] D2D transmission starts.
\end{enumerate}
\begin{figure*}[!t]
\centering
\includegraphics[width=6in]{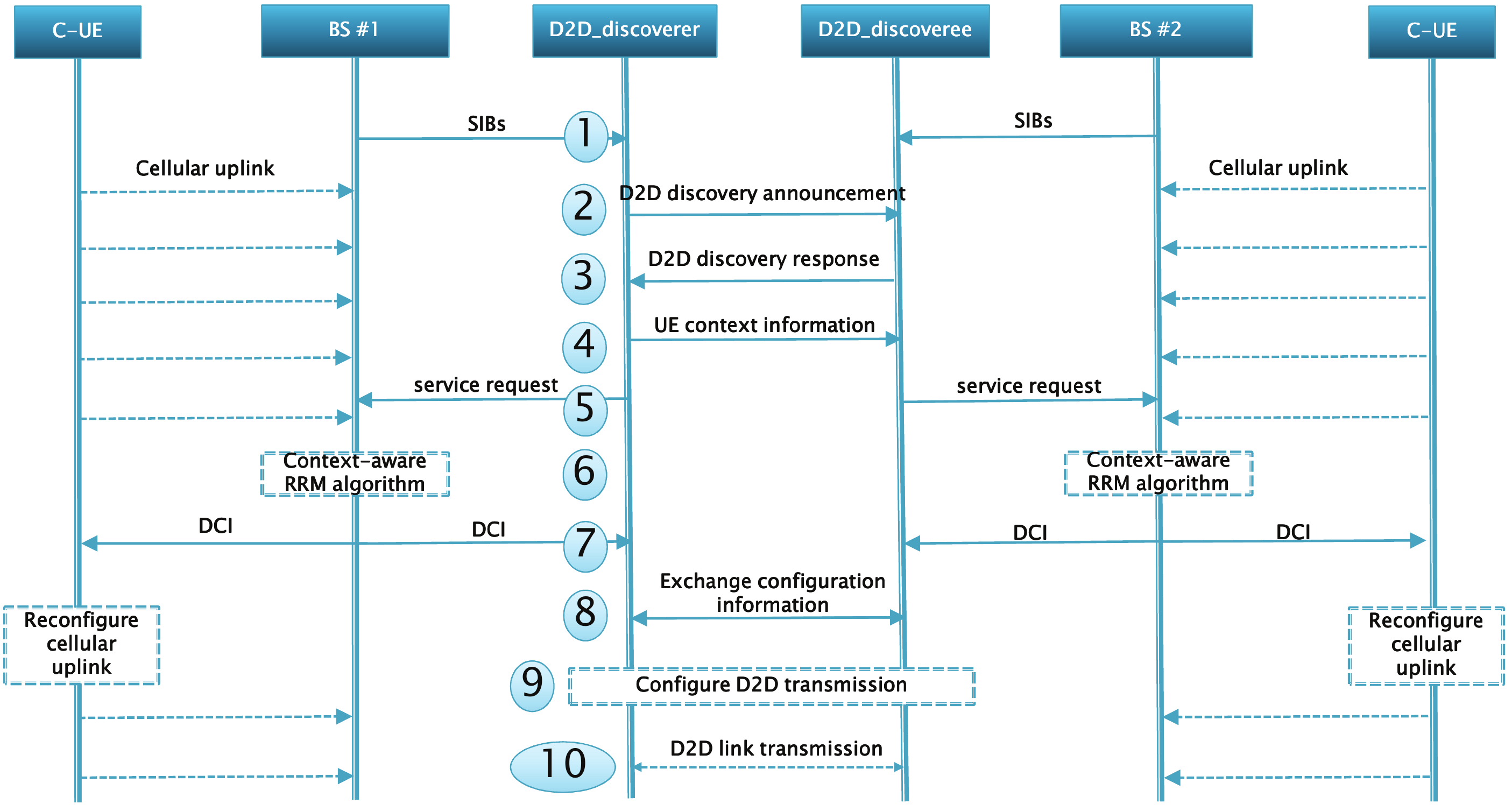}
\caption{Signaling scheme in multi cells}
\label{SSM}
\end{figure*}
Please note, in case D2D transmission is bi-directional, different resource can be configured for data transmission in different directions. Besides, the D2D discovery procedure happened in step 2) and step 3) does not require D2D ends to enter RRC connected state. In step 4), D2D discoverer needs to enter RRC connected state to send the D2D service request message. After sending the service request message, no matter whether D2D transmission is uni-directional or bi-directional, both the D2D transmitter and receiver need to listen to the DCI where D2D configuration information will be provided. Compared with the DCI used in legacy LTE-A network, the DCI for D2D operation is intended for one D2D pair instead of only one UE.\\
Moreover, in case if one D2D end has no cellular coverage, the same procedure can apply here. However, only the D2D user with cell coverage can send the D2D service request to the serving BS. After receiving the DCI, this user will forward this information to the other user who is out of cell coverage. \\
\subsection{Signaling scheme in multi cells}
Fig.~\ref{SSM} shows the proposed solution to enable D2D communication in a scenario where two D2D ends of one D2D link are served by two different BSs. As shown in Fig.~\ref{SSM}, the ten steps involved in the signaling diagram are illustrated with more details in the following, as
\begin{enumerate}
\item [1)]	Users exploiting D2D service receive SIBs broadcasted by their serving BSs and dedicated to D2D operation. In 3GPP, currently standardized SIBs provide information of not only resource pools used for transmission of D2D discovery and communication message in the serving cell, but also resource pools used in neighboring cells \cite{36331}. 
\item [2)]	One D2D discoverer UE and one D2D discoveree UE are paired by exploiting either ProSe direct discovery model A or discovery model B. Reference signals for channel estimation are also transmitted together with the discovery message. The D2D discovery message is sent over a resource indicated in the SIBs of the serving BS of D2D discoverer.
\item [3)]	Since the SIBs received from BS $\sharp$2 contain also information of resource pools used for D2D discovery and communication transmission in BS $\sharp$1, D2D discoveree can monitor over the resource where the discovery message was sent in previous step. If the D2D discovery request is accepted, the D2D discoveree replies an acknowledgement to the D2D discoverer, together with its own context information.
\item [4)]	When an acknowledgement message is detected by D2D discoverer, context information of the discoverer UE is transmitted towards the discoveree UE.
\item [5)]	Both discoverer and discoveree UEs send the service request with context information to their serving BSs.
\item [6)]	Both BSs analyse their gathered context information and perform its RRM algorithm. 
\item [7)]	For each of the BSs, if its serving D2D UE is allowed to reuse resource of uplink cellular users, a configuration message will be sent by the BS to the D2D end. The message contains information about which resource can be used for the D2D end to transmit D2D data. Meanwhile, the cellular link with the same resource may also need to be reconfigured.
\item [8)]	After receiving the configuration information from their serving BSs, two D2D ends will exchange their configuration information. Thus, both of them can be aware of the resources which they should listen to and receive D2D data from.
\item [9)]	Based on the acquired configuration information from step 7), each of D2D ends configures its transmission with the allocated resource. Based on the configuration information acquired from step 8), each of D2D ends configure its receiving with the corresponding resource.
\item [10)]	Bi-directional D2D transmission starts.
\end{enumerate}
\section{simulation assumptions and numerical results}\label{NR}
In order to evaluate system performance of the proposed D2D communication scheme, a system level simulator is developed. The applied simulator reflects real scenario, in order to obtain system performance as precise as possible. Detailed information regarding the simulation models and assumptions are captured in \cite{D21}. Here, we highlight the most important perspectives for our simulation in this section.
\subsection{Simulation models}
\subsubsection{Environmental}
\begin{figure}[!t]
\centering
\includegraphics[width=3.5in]{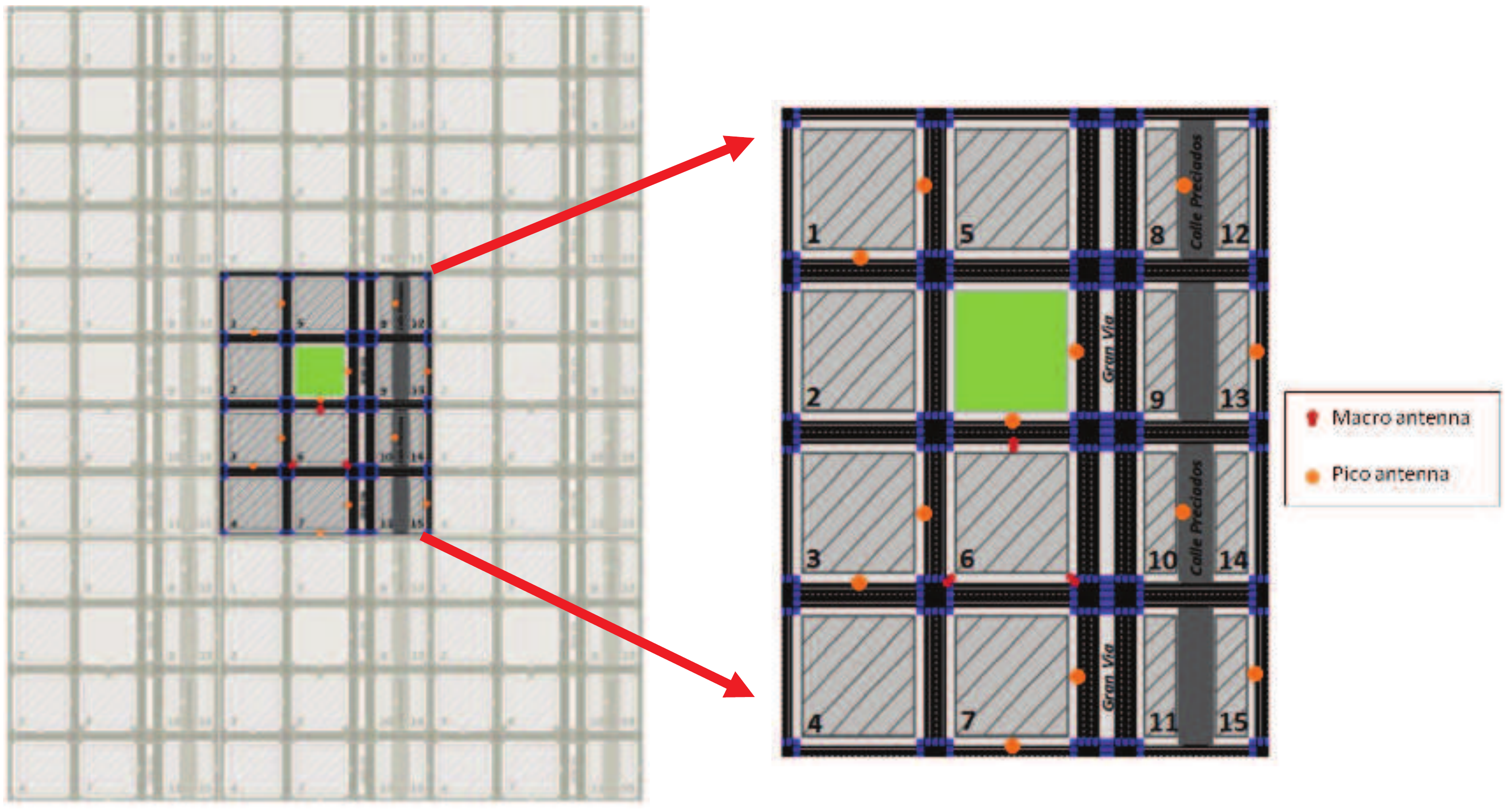}
\centering
\caption{Environment and deployment}
\label{envi}
\end{figure}
In this work, system performance of the proposed D2D communication is investigated in a dense urban environment. As shown in Fig.~\ref{envi}, a Madrid grid environmental model aligned with reality is used here \cite{D21}. In this model, an urban environment is depicted with 3D visualization where each grid composes of 15 buildings with different dimensions and one park. In order to avoid cell border effect, another eight replicas of Madrid grid are also placed but only users located in the central Madrid grid are inspected to derive system performance. Two dimensions for one Madrid grid are 387 m (east-west) and 552 m (south-north). Moreover, building heights are uniformly distributed between 8 and 15 floors with 3.5 m per floor.
\subsubsection{Deployment}
In Madrid grid, we inspect on users which are distributed outdoor with a user density of 1000 users per square kilometre. An isotropic antenna is installed on each user at 1.5 meter height with a maximal transmission power of 24 dBm. Besides, a heterogeneous network deployment with both macro and micro cells is used. For macro station, it operates in three cell sectors with carrier frequency of 800 MHz and directional antennas are positioned with 120 degree difference from each other in horizontal plane. At the same time, micro cells operate at central carrier frequency of 2.6GHz with two cell sectors per micro station where every micro cell points towards one side of the main street. Bandwidths of 10 MHz and 40 MHz are respectively used as cellular uplink resource for macro and micro cells.
\subsubsection{Traffic model}
Full buffer traffic model is used for both D2D and cellular links. Moreover, users are randomly selected to generate either cellular or D2D link traffic. 
\subsubsection{Channel model}
A 3D channel model proposed in \cite{D21} is used in this work which combines both a simplified ray-based approach and a stochastic and geometric approach. This model allows for a proper characterization of 3D environment in reality.
\subsubsection{Power control scheme and other simulation parameters}
Open loop power control is applied for both cellular and D2D links in order to compensate the propagation losses. One user device will set its transmit power to achieve certain SNR value at receiver end. Thus, interference is not considered in the open loop power control scheme.
\subsection{Numerical results and analysis}
\begin{figure}[!t]
\centering
\includegraphics[width=3.5in]{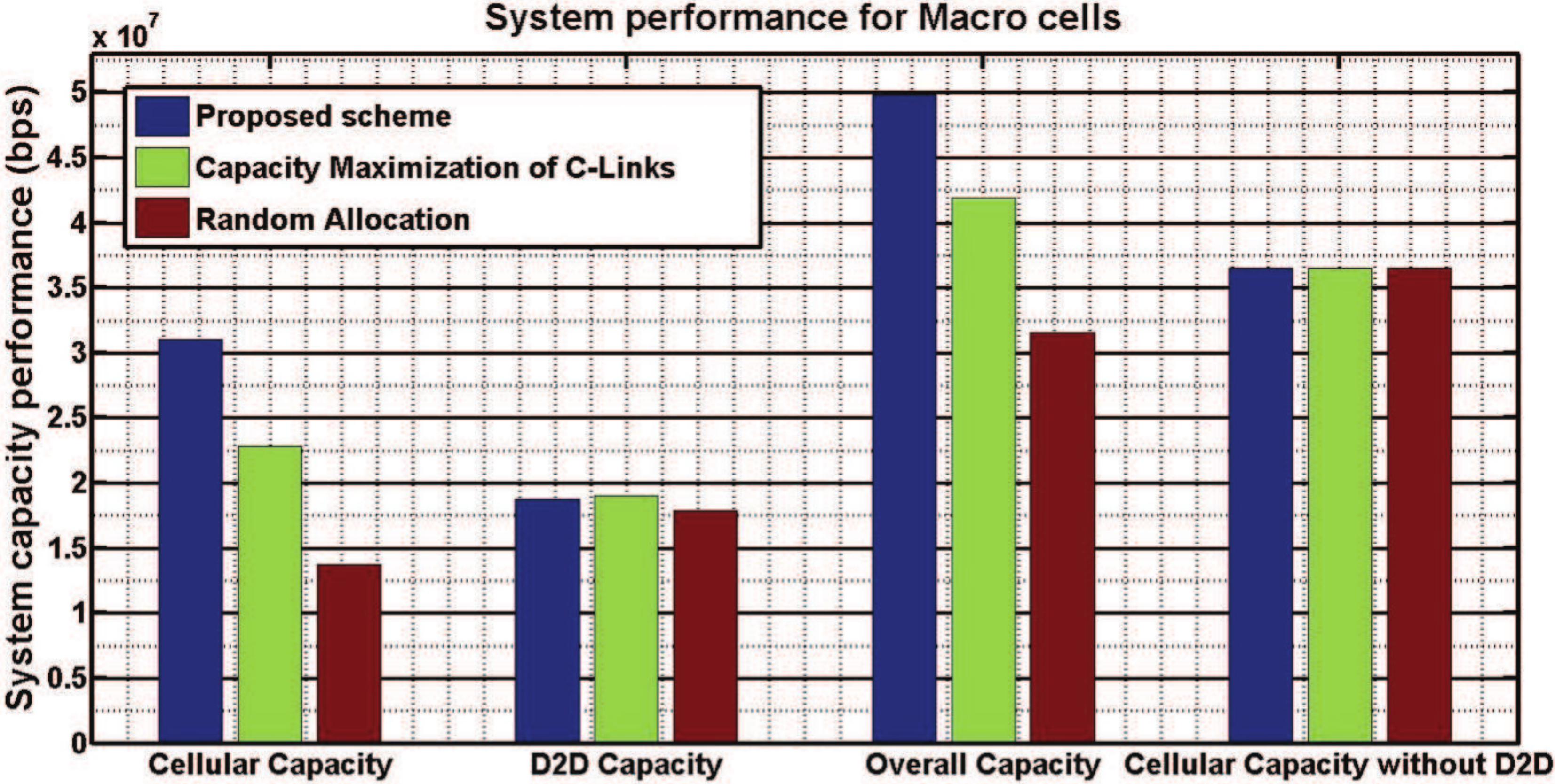}
\centering
\caption{System performance with only Macro cells (Scheme No.1 of D2D power control)}
\label{result1}
\end{figure}

\begin{figure}[!t]
\centering
\includegraphics[width=3.5in]{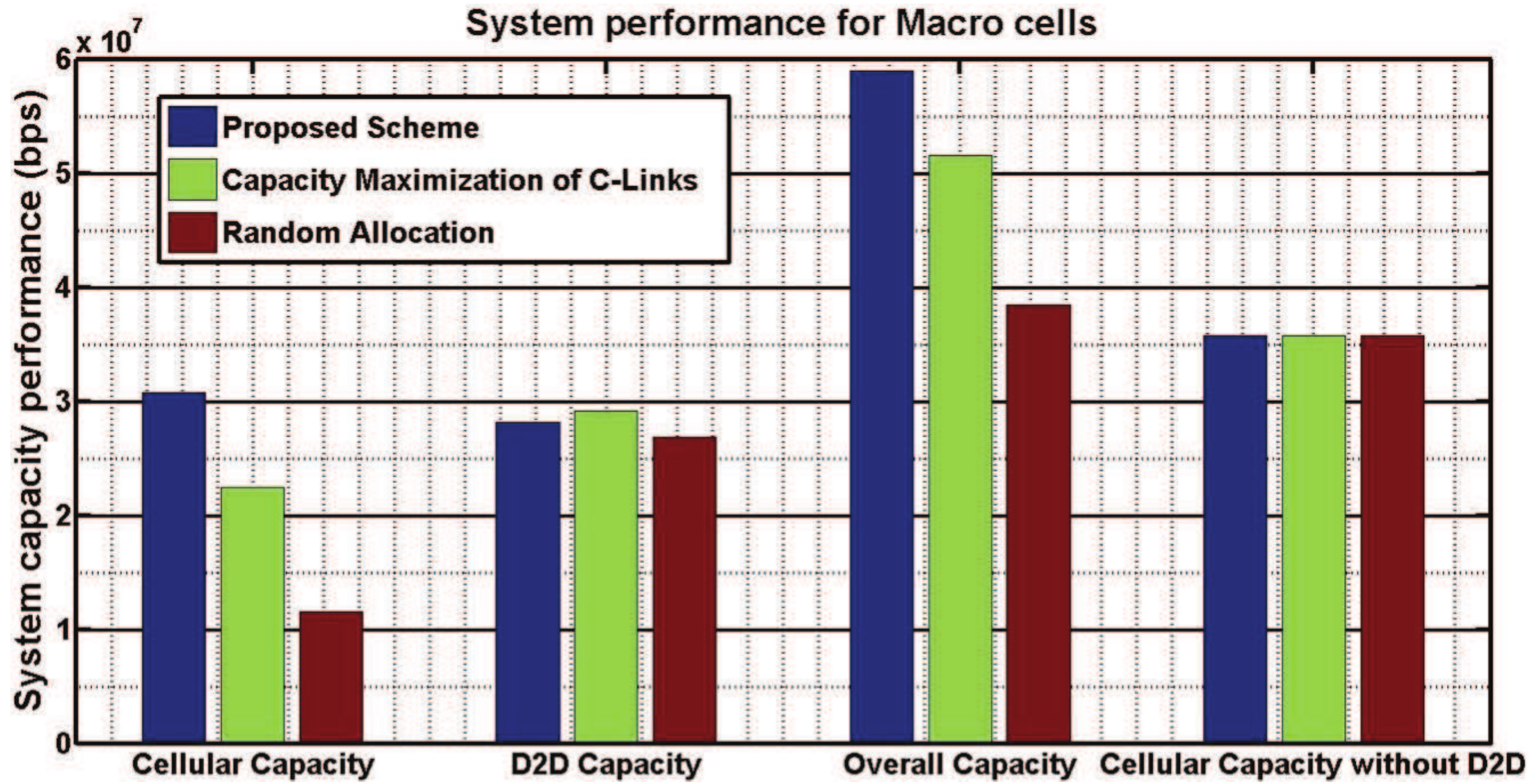}
\centering
\caption{System performance with only Macro cells (Scheme No.2 of D2D power control)}
\label{result2}
\end{figure}

\begin{figure}[!t]
\centering
\includegraphics[width=3.5in]{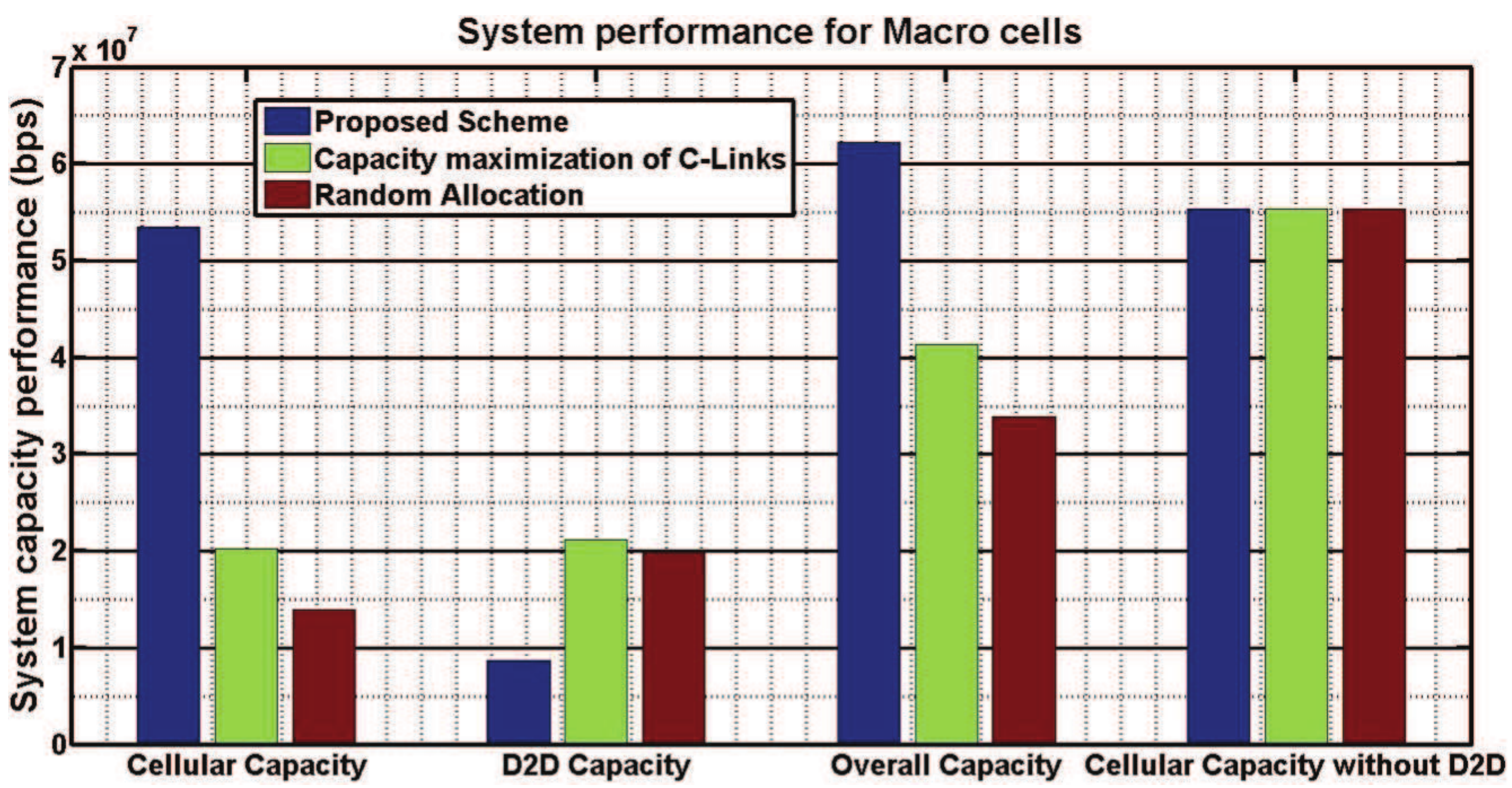}
\centering
\caption{System performance for macro cells in heterogeneous network}
\label{result3}
\end{figure}

\begin{figure}[!t]
\centering
\includegraphics[width=3.5in]{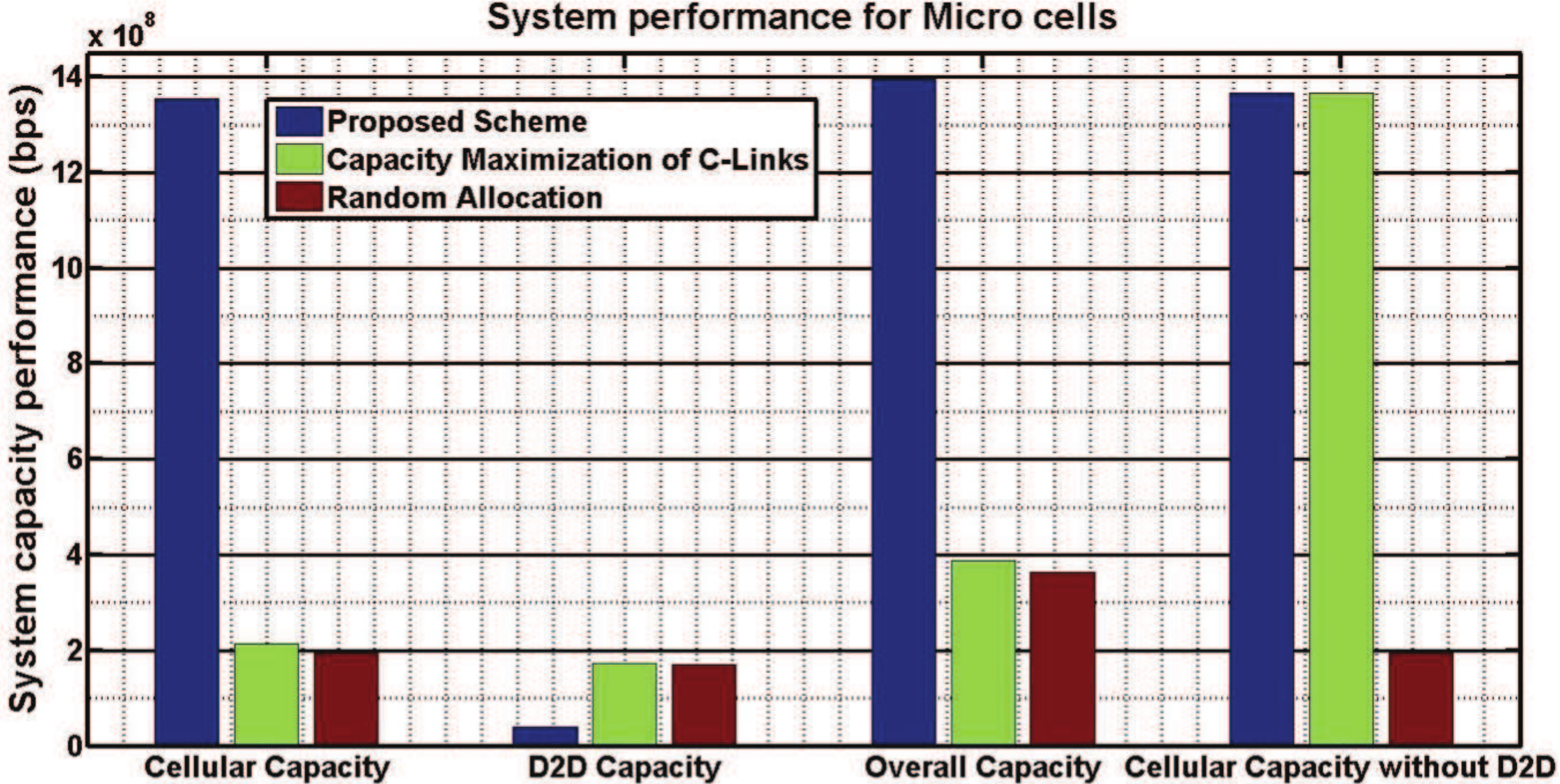}
\centering
\caption{System performance for micro cells in heterogeneous network}
\label{result4}
\end{figure}

In our work, we use a simple method to set the decision threshold for both cellular and D2D links to replace $SINR_{target}^{cell}$ and $\gamma_{(SINR_{target}^{D2D},d_m,\cdots)}$ in Eq.~(\ref{con2}) and Eq.~(\ref{DisRatio}) respectively, as
\begin{equation}
SINR_{(m,n)}^{cell}\geq SINR_n-\gamma_{cell},
\end{equation}
\begin{equation}
\frac{d_{(m,n)}}{d_m}\geq 1.
\end{equation}
$SINR_n$ represents the SINR value experienced by the $n$-th cellular link when no D2D link reuses the same resource. $\gamma_{cell}$ represents a deterioration offset allowed by RRM algorithm. In this work, the system is optimized w.r.t. the feasibility function defined in Eq.~(\ref{estF}) to enable as many D2D links as possible. The optimization problem can be efficiently solved by the algorithm proposed in \cite{my1}.\\
At first, we inspect on system performance of the proposed D2D communication with only macro station being deployed. In other words, the micro stations shown in Fig.~\ref{envi}  are shut off. The target SNR values for open loop power control are randomly generated in intervals [10dB,15dB] for cellular links and [0dB, 10dB] for D2D links. In Fig.~\ref{result1}, notation "Cellular Capacity" represents capacity of all macro cellular links, "D2D Capacity" represents capacity of all D2D links in reuse mode, "Overall Capacity" represents the overall capacity of both cellular and D2D links. As a comparison, the cellular capacity without D2D communication being used is also shown, with a notation of "Cellular Capacity without D2D". Moreover, the performance of another scheme denoted as "Capacity Maximization of C-Links" is also shown where capacity of cellular links are maximized by using the algorithm proposed in \cite{my2}. Last but not least, a random allocation of cellular resource to D2D links is also inspected and shown in this work, denoted as "Random Allocation". It is to be noted that all the resource of cellular links are reused by D2D links in the schemes "Capacity Maximization of C-Links" and "Random Allocation". However, the reuse of cellular resource is determined by the radio conditions of both cellular and D2D links in our proposed scheme. As it can be seen from Fig.~\ref{result1}, due to extra interference from D2D links in reuse mode, the capacity of cellular links in our proposed scheme has approximately decreased 16$\%$ compared with the case where no resource of cellular links is reused. However, thanks to the capacity gain brought by D2D communication, the overall capacity has been increased approximately with 35$\%$. Compared with the proposed scheme, the other two schemes have less gains in terms of the overall capacity. Since all cellular resource are reused in the scheme "Capacity Maximization of C-Links", there are more spectral resource available for D2D communication and thus the capacity of D2D links is higher than the proposed scheme. However, the additional D2D links introduce more interference to cellular links and thus a performance loss of cellular links can be seen in the figure. Moreover, it can also be seen from this figure that the random allocation of cellular resource to D2D links provides the least performance gain.\\
In Fig.~\ref{result2}, the same system setting is applied as in previous case except the target SNR value for open loop power control has now been increased to [7dB, 12dB] for D2D links. We can see from this figure that the cellular links experience a capacity deterioration of 15$\%$, which is same as in the previous case shown in Fig.~\ref{result1}.  This is due to the same interference control scheme for cellular links applied in RRM algorithm. However, higher SNR targets of D2D links introduce higher transmit power at D2D transmitters. Thus, D2D links contribute to higher capacity compared with the case shown in Fig.~\ref{result1} and the overall capacity has an increase of 60$\%$ compared with legacy cellular network where no D2D communication is allowed.\\
Micro cells shown in Fig.~\ref{envi} are now turned on and system performance of the heterogeneous network is inspected. The D2D target SNR values are set in [0dB,10dB], same setting as shown in Fig.~\ref{result1}. Fig.~\ref{result3} and Fig.~\ref{result4} demonstrate the system performance of macro cells and micro cells, respectively. As can be seen, the cell capacity without D2D links is higher than the previous cases. This is due to the presence of micro cells, and users experiencing low SINR values from macro cell will now be served by micro cells. Thus, users served by macro cells experience better channel situation compared with the previous cases. In other words, users are now served by macro cells if they locate closely to macro BS, otherwise they would be served more likely by micro cells. Therefore, if a D2D user in macro cell coverage is allocated to reuse resource of one cellular link, a higher interference power level for cellular link will be introduced by the D2D transmission. Thus, less D2D links are allowed to reuse cellular resource. Now, we can see that overall capacity has an improvement of 13$\%$ in the proposed scheme which is less than the case shown in Fig.~\ref{result1}. On the other side, since less D2D links are assigned with resource, less cellular links are deteriorated. Therefore, there is only a very small amount of capacity decrease for cellular links, compared with the legacy system. Moreover, due to the deployment of micro cells, the coverage of one cell is much smaller than the case shown in Fig.~\ref{result1}. Therefore, the mutual interference in between cellular and D2D links can be so severe that the impact of the deterioration of radio conditions can not be compensated by the exploitation of D2D communication and thus needs to be controlled more properly. This is the reason why the overall capacities of the other two schemes are even lower than the case where no D2D communication is applied. In Fig.~\ref{result4}, capacity of micro cells is demonstrated. Due to short coverage of micro cells, very few cellular resource can be reused by D2D links and a capacity increase of 4$\%$ is introduced by using our scheme.
\section{Conclusion}\label{con}
In this paper, we proposed one context-aware RRM algorithm which allows for D2D communication in reuse mode with a reasonable signaling effort. Signaling schemes to support the proposed RRM algorithm are also given for both single cell and multi cell scenarios. Numerical result shows our context-aware RRM algorithms with support of the proposed signaling schemes yield a good system performance in terms of overall system capacity.   
\section{Acknowledgement}
Part of this work has been performed in the framework of H2020 project METIS-II, which is funded by the European Union. The views expressed are those of the authors and do 
not  necessarily  represent  the  project. The consortium is not liable for any use that may be made of any of the information contained therein.


%
%



%

\end{document}